\documentstyle{article}
\begin{document}

\begin{center}
{\bf \large \bf Faddeev approach to the octet and decuplet baryons}\\
\vskip 0.2cm
C. Hanhart and S. Krewald\\
\vskip 0.2cm

 Institut f\"{u}r Kernphysik,\\
 Forschungszentrum J\"ulich GmbH,\\
 D--52425 J\"ulich, Germany
 \end{center}

\begin{abstract}
A relativistic Faddeev model for the baryon octet is
 extended to treat the baryon decuplet.
We find that after determining the model parameters in the mesonic
sector the masses of both nucleon and delta deviate by less than 5\%
from the experimental data and show only a very weak dependence
on the constituent quark mass.
\end{abstract}



Recently relativistic Faddeev quark--models have been developed for
the nucleon [1-6].  So far, the Faddeev approach was applied to the
baryon octet only.  In this communication, we extend the Faddeev
approach to the baryon decuplet as well.  The aim of our investigation
is to check whether a consistent description of mesonic and baryonic
properties is possible within this framework.

The relativistic Faddeev models are especially suited to investigate
diquarks [7-9] inside the baryons, since diquark substructures appear
naturally in these approaches. The amount of correlation is determined
by the baryon properties.  Therefore the appearance of bound diquark
states is not essential.

Quark models of the nucleon have to respect chiral symmetry. The
Nambu--Jona-Lasinio (NJL) model \cite{njl} provides a particularly
simple implementation of chiral symmetry as it is reviewed in
refs. \cite{weise,kle}. As most relativistic Faddeev models we
therefore start from a NJL-type Lagrangian:

\begin{equation}
{\cal L}=\bar q (i \partial \!\!\! / -m_0)q
 -G_1(\bar{q}\tilde{\Lambda}_1^{\mu k}q)
                  (\bar{q}\tilde{\Lambda}_{\mu1}^{k}q)
                        + G_2(\bar{q}\tilde{\Lambda}_2^{\mu k}q)
                  (\bar{q}\tilde{\Lambda}_{\mu2}^{k}q) \ ,
\end{equation}

where $m_0=diag(m_0^u,m_0^d,m_0^s)$,
$
\tilde{\Lambda}^{\mu}_1 =
\gamma^{\mu}\otimes I_{f}\otimes\lambda_{c}^{k}$
and
$
\tilde{\Lambda}^{\mu}_2
=\gamma^{\mu}\gamma_5 \otimes I_{f}\otimes\lambda_{c}^{k} \ ,
$
$\lambda_{c}^{k}$ being the Gell-Mann matrices.

Using the Fierz transformation the interaction can be expressed in
terms of a particle--antiparticle interaction with suitable colour
channels

\begin{equation}
\label{ladi2}
{\cal L}_{I}^M=\frac{g_1^a}{2}(\bar{q}\Lambda^{[1]}q)
                           (\bar q\Lambda^{[1]}q)+
             (colour-octet-terms) \ ,
\end{equation}

where
$
 \Lambda^{[1]}=I_{c}^{\gamma}\otimes\lambda^{k}_{f}\otimes\Gamma^{a}
$
and a particle--particle interaction

\begin{equation}
{\cal L}_{I}^D
= -\frac{g_2^a}{2}(\bar{q}\Lambda^{[\bar{3}]}C\bar{q}^{T})
                           (q^{T}C^{-1}\Lambda^{[\bar{3}]}q)
+  (colour-sextet-terms) \ ,
\end{equation}

where $ \Lambda^{[\bar{3}]}
=t_{c}^{\gamma}\otimes\lambda^{k}_{f}\otimes\Gamma^{a} \ , $ with
$\Gamma^a_s \in \{I,i\gamma _5 \}$ and $\Gamma^a_v \in \{ i\gamma
^{\mu},i \gamma ^{\mu} \gamma _5 \}$.  The color antitriplet is
generated by $t_{\alpha \beta}^{\gamma}
=i\sqrt{\frac{3}{2}}\epsilon_{\alpha \beta \gamma}$.  The generators
of the flavor group $U(3)_F$ are denoted $\lambda_F^A$ by $A \in
\{0,..,8\}$.  The coupling constants $g_1$ and $g_2$ are functions of
$G_1$ and $G_2$: $g_1^s=\frac{16}{9}(G_1+G_2)$,
$g^v_i=\frac{1}{2}g^s_i$ and $g_1^a=\frac{1}{2}g_2^a$.  For $G_2=0$
the above Lagrangian can be viewed as the Fierz transformation of the
point like interaction of two color octet currents \cite{rein}.  The
investigation of meson properties in the NJL model, however, proved
that the scalar and vector coupling constants have to be varied
independently \cite{ebert}.  The recent work of Ishii et
al. \cite{ishii2} has shown that the scaling factor
$\alpha=\frac{g^a_1}{g^a_2}=\frac{1}{2}$ mentioned above produces a
proton mass quite close to the experimental one provided the axial
vector diquark is considered in addition to the scalar diquark.

  We take into account only the two low lying diquarks
namely the scalar one ($\Gamma_s = i\gamma_5$) and the axial vector
one ($\Gamma_v^{\mu} = i\gamma^{\mu}$).


The two--body T-matrices are calculated in the ladder approximation.
 For the quark--quark sector this yields:
\begin{equation}
\label{dipo}
T(q^2)={\tilde T}(q^2)[\lambda^A_F
\otimes t^\gamma \otimes C^{-1}\Gamma^\mu]
{\cal O}_{\mu \nu}
[\lambda^A_F \otimes t^\gamma \otimes \Gamma^\nu C]
\end{equation}
with ${\tilde T}(q^2)=\left(\frac{1}{2g_2}-\frac{1}{2}J_D(q^2)
 \right)^{-1}$, ${\cal O}_{\mu \nu}$ as the vector structure of the
 two--body Greens function (${\cal O}=I$ for the scalar diquark,
 ${\cal O}_{\mu \nu}=g_{\mu \nu}-q_\mu q_\nu /q^2$ for the axial
 vector diquark) and
\begin{equation}
\label{j}
J_D(q^2)=-i \int \frac{d_4k}{(2\pi )^4}tr(\Lambda^{[\bar{3}]}
S_F(-q_1) \Lambda^{[\bar{3}]} S_F(q_2)) \ .
\end{equation}
where $q=q_1+q_2$ and we used $C^{-1}S^T_F(q)C=S_F(-q)$.  Since the
quarks are treated as identical particles the Pauli principle requires
the diquark vertices to be antisymmetric \cite{weise}.  For the mesons
we get ${\tilde T}_M(q^2)= \left( \frac{1}{g_1}-J_M(q^2) \right)^{-1}$
and
\begin{equation}
\label{m}
J_M(q^2)= i \int \frac{d_4k}{(2\pi )^4}tr(\Lambda^{[1]}  S_F(-q_1)
\Lambda^{[1]} S_F(q_2)) \ .
\end{equation}
The Lagrangian leads to both bound diquarks and bound mesons, although
the corresponding terms appear with different sign in
eqn. (\ref{ladi2}). This difference is compensated by an additional
minus sign in the meson polarization function (cf. eqn. (\ref{m}) and
eqn. (\ref{j})) created by its closed fermion loop. The two--body
T-matrices are diagonal not only in color and flavor space, but also
do not mix channels due to the traces appearing in eqns. (\ref{j}) and
(\ref{m}).

A detailed derivation of the relativistic Faddeev equations is given
in refs.
\cite{rein,huang}.
One has to use the same kind of approximation as has already been used
in the meson case.  The essential one is to neglect the three--body
irreducible graphs.  This reduces the three--body problem to an
effective two--body Bethe--Salpeter equation with an energy dependent
interaction.

The Faddeev amplitude therefore reads
\begin{eqnarray}
\label{an}
\nonumber
{\cal T}(p_1,p_2,p_3)_{\alpha \beta \gamma}&=&
\phantom{+}(C^{-1}t^{\gamma}\lambda^A \gamma_5)_{\alpha \beta}
{\tilde{T}}(p_1+p_2)^{AA} \Psi(P,p_3)^{A}_{\gamma} \\
\nonumber
& &+(C^{-1}t^{\gamma} \lambda^S \gamma^{\mu})_{\alpha \beta}
{\tilde{T}}(p_1+p_2)^{SS}_{\mu \nu} \Psi(P,p_3)^{S \ \nu}_{\gamma}\\
& &+(cyclic) \ ,
\end{eqnarray}
with $\lambda^{A,S}$ being the antisymmetric and symmetric Gell-Mann
matrices respectively.  Here $cyclic$ means cyclic permutation of all
types of indices at once.  The amplitudes $\Psi$ are the solutions of
the following integral equation:
\begin{equation}
\label{psi}
 \Psi(P,q)^A_{(\mu)}=i \int \frac{d_4k}{(2\pi )^4}
 L(k_1,k_2,P)^{AC}_{(\mu \tau)} \Psi(P,P-k)^{C \ (\tau)} \ ,
\end{equation}
where $k_1=q-k$, $k_2=k$ and
\begin{equation}
\label{l}
 L(q,p,l)^{AC}_{(\mu \tau)}=C^{-1}\Lambda^{B \
(\nu)}S_F(q)\Lambda^{A}_{(\mu)}CS_F(p)^T{\tilde{T}}(l-p)^{BC}_{(\nu
\tau)} \ .
\end{equation}
The brackets around the Lorentz indices indicate
that they only appear, if the axial vector diquark is involved.

As was observed in ref. \cite{buck}  the color part of $L$ can
be written as a sum of two projectors
$(t^B t^A=-3P^{[1]}+\frac{3}{2}P^{[8]})$.
This means that no mixing between the color singlet and the
color octet occurs. Due to their different signs only one
of the possible color multiplets leads to bound states. In
our case this is the color singlet. Therefore in the
following we only have to consider color singlet states.
As all the indices are fixed here, the color indices are
omitted in the following.

The next step is to project out from eqn. (\ref{an}) the flavor
multiplets assuming exact SU(3)$_F$.  The suitable projectors can in
general be constructed by the tools provided in ref. \cite{dS}.  The
projector $P^{[10]}$ on the decuplet symmetrizes all indices, the
singlet projector $P^{[1]}$ antisymmetrizes and $P^{[8]}=
id-P^{[10]}-P^{[1]}$.  The subspace of the octet obtained by this
construction contains both types of octets appearing in the
decomposition of the flavor content of the three quarks.  At this
stage it is crucial to observe that the amplitudes $\Psi$ only appear
in certain combinations in the equations derived by the above
procedure when the explicit representation of the Gell-Mann matrices
is introduced.  These linear combinations are respected by the flavor
content of $L$. Therefore this construction provides the basis that
block diagonalizes $L$ (table (\ref{8ta8})).  Since we know the flavor
structure of every block it is possible to uniquely identify the
particle described by a special subspace.

\begin{table}
\caption{Quark content and basis for baryon amplitudes.}
\renewcommand{\baselinestretch}{1}
\label{8ta8}
\begin{center}
\begin{tabular}{|c|c|c|}
\hline
 & $ ^{[8]} \Phi ^{a \nu} $
& $^{[8]}\Phi^s$ \\
\hline
\hline
p  \ (uud) & $\frac{1}{\sqrt{3}}\left(
 \Psi ^{a_{(ud)}\nu}_{u}-\sqrt{2}\Psi ^{a_{(uu)}\nu}_{d} \right) $ &
$ i\Psi ^{s_{(ud)}}_{u} $\\
\hline
n \ (udd) & $\frac{1}{\sqrt{3}}\left(
 \Psi ^{a_{(ud)}\nu}_{d}-\sqrt{2}\Psi ^{a_{(dd)}\nu}_{u} \right) $ &
$ i\Psi ^{s_{(ud)}}_{d} $\\
\hline
\hline
$\Sigma^{+}$ \ (uus) & $\frac{1}{\sqrt{3}}\left(
 \Psi ^{a_{(us)}\nu}_{u}-\sqrt{2}\Psi ^{a_{(uu)}\nu}_{s} \right) $ &
$ i\Psi ^{s_{(us)}}_{u} $\\
\hline
$\Sigma^{o}$ \ (uds) & $\frac{1}{\sqrt{6}} \left(
 \Psi ^{a_{(ds)}\nu}_{u}+\Psi ^{a_{(us)}\nu}_{d}-2\Psi
^{a_{(ud)}\nu}_{s} \right) $ &
$\frac{i}{\sqrt{2}}\left(  \Psi ^{s_{(us)}}_{d}+
\Psi ^{s_{(ds)}}_{u} \right) $\\
\hline
$\Sigma^{-}$ \ (dds) & $\frac{1}{\sqrt{3}}\left(
\Psi ^{a_{(ds)}\nu}_{d}-\sqrt{2}\Psi ^{a_{(dd)}\nu}_{s} \right)$ &
$ i\Psi ^{s_{(ds)}}_{d} $\\
\hline
\hline
$\Xi^{o}$ \ (uss) & $\frac{1}{\sqrt{3}}\left(
 \Psi ^{a_{(us)}\nu}_{s}-\sqrt{2}\Psi ^{a_{(ss)}\nu}_{u} \right) $ &
$ i\Psi ^{s_{(us)}}_{s} $\\
\hline
$\Xi^{-}$ \ (dss) & $\frac{1}{\sqrt{3}}\left(
\Psi ^{a_{(ds)}\nu}_{s}-\sqrt{2}\Psi ^{a_{(ss)}\nu}_{d}\right) $ &
$ i\Psi ^{s_{(ds)}}_{s} $\\
\hline
\hline
$\Lambda$ \ (uds) & $\frac{1}{\sqrt{2}}\left(
\Psi ^{a_{(ds)}\nu}_{u}-\Psi ^{a_{(us)}\nu}_{d} \right)$ &
$\frac{i}{\sqrt{6}}  \left( 2\Psi ^{s_{(ud)}}_{s}-
\Psi ^{s_{(ds)}}_{u}+\Psi ^{s_{(us)}}_{d} \right) $\\
\hline
\multicolumn{3}{c}{ } \\
\hline
 & $ ^{[10]} \Phi ^{a \nu} $ &
 \\
\hline
\hline
$\Delta^{++}$ \ (uuu) & $\Psi ^{a_{(uu)}\nu}_{u} $ &\\
\hline
$\Delta^{+}$ \ (uud) & $\frac{1}{\sqrt{3}}\left(
\Psi ^{a_{(uu)}\nu}_{d}+\sqrt{2}\Psi ^{a_{(ud)}\nu}_{u} \right) $&\\
\hline
$\Delta^{o}$ \ (udd) & $\frac{1}{\sqrt{3}}\left(
\Psi ^{a_{(dd)}\nu}_{u}+\sqrt{2}\Psi ^{a_{(ud)}\nu}_{d} \right) $&\\
\hline
$\Delta^{-}$ \ (ddd) & $\Psi ^{a_{(dd)}\nu}_{d} $& \\
\hline
\hline
$\Sigma^{+*}$ \ (uus) & $\frac{1}{\sqrt{3}}\left(
 \Psi ^{a_{(uu)}\nu}_{s}+\sqrt{2}\Psi ^{a_{(us)}\nu}_{u} \right) $
& \\
\hline
$\Sigma^{o*}$ \ (uds) & $\frac{1}{\sqrt{3}}\left(
\Psi ^{a_{(ud)}\nu}_{s}+\Psi ^{a_{(ds)}\nu}_{u}+
\Psi ^{(su)_{s}\nu}_{d} \right) $& \\
\hline
$\Sigma^{-*}$ \ (dds) & $\frac{1}{\sqrt{3}}\left(
 \Psi ^{a_{(dd)}\nu}_{s}+\sqrt{2}\Psi ^{a_{(ds)}\nu}_{d}
 \right) $ & \\
\hline
\hline
$\Xi^{o*}$ \ (uss) & $\frac{1}{\sqrt{3}}\left(
\Psi ^{a_{(ss)}\nu}_{u}+\sqrt{2}\Psi ^{a_{(su)}\nu}_{s} \right) $& \\
\hline
$\Xi^{-*}$ \ (dss) & $\frac{1}{\sqrt{3}}\left(
 \Psi ^{a_{(ss)}\nu}_{d}+\sqrt{2}\Psi ^{a_{(ds)}\nu}_{s} \right) $&\\
\hline
\hline
$\Omega^{-}$ \ (sss) & $\Psi ^{a_{(ss)}\nu}_{s} $&\\
\hline
\multicolumn{3}{c}{ } \\
\hline
& &
 $^{[1]}\Phi^a$ \\
\hline
\hline
-  (uds) &&
$ \frac{i}{\sqrt{3}}\left(-\Psi ^{s_{(ud)}}_{s}-
\Psi ^{s_{(ds)}}_{u}+\Psi ^{s_{(su)}}_{d}\right) $\\
\hline
\end{tabular}
\end{center}
\end{table}

Within this basis we obtain the following set of equations:\\
for the singlet:
\begin{equation}
\label{sin}
^{[1]}\Phi^s = -2A^{(ss) \  [1]}\Phi^s \ ,
\end{equation}
for the octet:
\begin{eqnarray}
\nonumber
\label{o}
 ^{[8]} \Phi ^{a \ \mu}&=&-A^{(aa) \mu
 \ [8]}_{\ \ \ \ \ \nu} \Phi ^{a \ \nu}
-\sqrt{3}A^{(as) \mu \ [8]}\Phi^s\\
^{[8]}\Phi^s&=&-\sqrt{3}A^{(sa) \ [8]}
_{\ \ \ \nu} \Phi ^{a \ \nu}+A^{(ss) \ [8]}\Phi^s \ ,
\end{eqnarray}
and for the decuplet:
\begin{equation}
\label{de}
 ^{[10]} \Phi ^{\mu}=2A^{(aa) \mu \ [10]}_{\ \ \ \ \ \nu} \Phi ^{\nu}
\ ,
\end{equation}
where $A$ is a linear operator defined in the following way
\begin{equation}
\label{opa}
(A^{AB} \Phi^B)(P,q) := i \int \frac{d_4k}{(2\pi )^4}
 L_D(q-k,k,P)^{AB}_{(\mu \tau)} \Phi(P,P-k)^{B \ (\tau)} \ .
\end{equation}
Here $L_D$ is the Dirac part of
 the matrix $L$ defined in (\ref{l}) and $a$ and $s$ stand for
a suitable axial vector or scalar diquark respectively.

Eqns. (\ref{sin})--(\ref{de}) now in principle can be solved
numerically.  Given the relatively good performance of the static
approximation \cite{ishii,ishii2}, we simplify the numerical work by
replacing the propagator of the exchanged quark by $S_F(q) \rightarrow
-\frac{1}{M}$ following Buck et al. \cite{buck}. The scattering kernel
$L$ is therefore independent of the outgoing diquark momentum $q$
which totally disappears from eqn. (\ref{psi}). Consequently  the
amplitudes $\Phi$ only depend on the baryon momentum $P$. $A$ is no
longer an integral operator but a simple matrix.  The remaining
integral has to be regularized. We used a sharp Euclidean cut off
$\Lambda$.  The treatment of the singularities in the integral
required to perform a Wick rotation is discussed in \cite{ishii}.
To ensure good spin of the solutions we used the spin projectors given
in reference \cite{niew}.

The equations (\ref{sin})--(\ref{de}) are valid only in the limit of
SU(3)$_F$ symmetry. If one introduces different quark masses, the
operator $A$ defined in equation (\ref{opa}) receives additional
contributions which couple the equations for octet and decuplet.  As
long as we keep exact SU(2)$_F$ the equation for the $\Lambda$ is
mixed with the flavour singlet only and there is no mixing for the
nucleon and the delta.  Using the spin projectors for
spin--$\frac{3}{2}$ and spin--$\frac{1}{2}$ we get two sets of coupled
equations. The solutions can be naturally identified with the
observables.

The model contains 5 parameters: 2 current masses ($m_0^u=m_0^d, \
 m_0^s$), the cut off ($\Lambda$) and 2 coupling constants ($g_1^s, \
 g_1^v$).  On the other hand we have 14 observables:

$\bullet$ 4 masses from the baryon decuplet
($M_\Delta, \ M_{\Sigma^*}, \ M_{\Xi^*}, \ M_{\Omega^-}$),

$\bullet$ 4 masses from the baryon octet
 ($M_N, \ M_\Sigma, \ M_\Lambda, \ M_\Xi$),

$\bullet$ 3 masses from the nonet of the vector mesons
 ($m_\rho=m_\omega, \ m_{K^*}, \ m_\Phi$), where the $\Phi$ is treated
 as pure $s \bar s$ state,

$\bullet$ 2 masses form the nonet of the pseudoscalar mesons ($m_\pi,
\ m_{K}$)(To calculate the mass of the $\eta`$--meson one would have to
include the t'Hooft interaction \cite{weise}) and

$\bullet$ the pion decay constant $f_\pi$ using the
Gell-Mann--Oaks--Renner relation
\cite{gor}.

For a given Cut--off we can therefore fix all 4 parameters using only
mesonic data.  We investigate the behavior of the spectrum in
dependence of the constituent mass of the up quark. The cut off is
determined by the NJL gap eqn. \cite{njl}. Note that we have the
constraint of $m_u^* \geq 411$ MeV in order to bind the
$\Delta_{33}(1232)$.

As a check of our formalism, we switched off the vector coupling in
 the particle--particle channel ($g_2^v=0$) and reproduced the results
 of Buck et al. \cite{buck}, provided the scalar coupling is increased
 to reproduce the nucleon mass.  In this case a strongly bound scalar
 diquark is found ($m_{s_{(ud)}}=605$ MeV with a binding energy of
 $E_{s_{(ud)}}=295$ MeV for $m_u^*=450$ MeV).  We confirm the finding
 of \cite{ishii2}, that if both scalar and axial vector diquark
 channels are considered simultanously, the binding energy of the
 three--body system is increased.  Therefore a less bound scalar
 diquark is required for the binding of the nucleon.  A similar result
 was found in a recent analysis of lattice gauge calculations
 \cite{lein}.

For constituent quark masses below 440 MeV we had no bound axial
vector diquarks, but for larger masses binding occured.

\begin{figure}
\vspace{14cm}
\includegraphics{bletter.ps}
\caption{{The deviation of the numerical results from the experimental
data in percent.}}
\label{t3}
\end{figure}

The deviation of the masses of decuplet and octet baryons calculated
in our model from the experimental values is shown in figure \ref{t3}
as a function of the quark mass. The masses of both nucleon and
$\Delta_{33}(1232)$ are reproduced within 5 \% for quark masses
between 411 MeV and 800 MeV. With increasing quark masses the coupling
constants are adjusted so that the meson masses remain constant. This
suffices to produce both a nucleon and a delta mass which show only
weak dependence on the constituent quark mass.

In the solitonic approach to the baryons of Weigel et
al. \cite{weigel}, the best mass splittings were obtained with a
constituent quark mass of 390 MeV. In our model the fit quality is
best for the smallest quark mass of 411 MeV, but unfortunately we
cannot further decrease the quark mass, since our model lacks a
confinement mechanism.

For a constituent quark mass of 411 MeV, the masses of the strange
baryons deviate from the experimental ones by approximately 15 \%
(table (\ref{ergI})) and one can not improve the agreement with the
data by increasing the quark masses.  We checked that treating the
scaling factor $\alpha=\frac{g_1^a}{g_2^a}$ as a free parameter
reduces the maximal deviation from the experimental masses to 7 \%.

\begin{table}
\caption{The numerical results for the baryon masses for a fixed
 constituent mass $m_u^*=411$ MeV. Deviations from the experimental
 values in percent are added in brackets. Data used to fix the
 parameters are stared (*). For the constituent mass choosen the axial
 vector diquarks remained unbound and are therefore not mentioned.}
\baselineskip=1cm
\renewcommand{\baselinestretch}{1}
\label{ergI}
\begin{center}
\baselineskip=1cm
\begin{tabular}{|c||r|r|}
\hline
&M [MeV]& $E^b$ [MeV] \\
\hline
\hline
\multicolumn{3}{|c|}{quarks} \\
\hline
$m_0^u$ & 8.4 \ & --  \ \\
$m_0^s$ &204 \  & -- \   \\
$m_s^*$ & 617 \  & -- \  \\
\hline
\hline
\multicolumn{3}{|c|}{mesons} \\
\hline
$f_{\pi}$& $93^* \ \phantom{(3)}$ &-- \   \\
$\pi$ & $135^* \ \phantom{(3)}$  & 687 \  \\
$K$ & $495^* \ \phantom{(3)}$   & 533  \   \\
$\rho$ & $770^* \ \phantom{(3)}$ & 52  \   \\
$K^*$ & $938$ \ (5)    & 105  \     \\
$\Phi$ & 1096 \ (7)    & 138 \  \\
\hline
\hline
\multicolumn{3}{|c|}{diquarks} \\
\hline
$s_{(ud)}$ & 710 \ \phantom{(3)}  & 112  \  \\
$s_{(us)}$ & 899 \ \phantom{(3)}  & 129  \   \\
\hline
\hline
\multicolumn{3}{|c|}{baryons} \\
\hline
$N$ & 901 \ (4)& 220 \  \\
$\Sigma$ & 1081 \ (9) & 229  \ \\
$\Lambda$ & 941 (16) & 369 \ \\
$\Xi$ & 1114 (15) & 402  \ \\
\hline
$\Delta$ & 1212 \ (2) & 21 \ \\
$\Sigma^*$ & 1401 \ (1) & 38 \  \\
$\Xi^*$ & 1597 \ (4) & 48 \  \\
$\Omega^-$ & 1801  \ (8) & 50 \ \\
\hline
\end{tabular}
\end{center}
\end{table}

The fact that the strange baryons deviate more strongly from the data
than the proton and the delta may be taken as a hint that the t`Hooft
interaction should not be neglected. The t`Hooft interaction affects
the scalar diquarks only and therefore only the octet baryons.


To summarize, we have shown that the Faddeev approach to the baryonic
sector of the Nambu--Jona-Lasinio model is able to provide a
consistent description of both octet and decuplet baryons.

Thanks is given to I. Afnan, B. Gibson, H. Haberzettl, N. Nakayama and
J. Speth for useful discussions.

\end{document}